\documentclass[a4paper,11pt]{article}

\usepackage[utf8]{inputenc}   
\usepackage{amsmath}          
\usepackage{amssymb}          
\usepackage{graphicx}         
\usepackage{color}
\usepackage{hyperref}         
\usepackage{geometry}         
\usepackage{natbib}           
\usepackage{authblk}          
\usepackage{appendix}
\usepackage{multirow}

\title{What causes long outbursts of neutron star low-mass X-ray binaries?}
\author[1]{Sayantan Bhattacharya\thanks{
sayantan34@gmail.com}}
\author[1]{Sudip Bhattacharyya}
\affil[1]{Department of Astronomy and Astrophysics, Tata Institute of Fundamental Research, 1 Homi Bhabha Road, Colaba, Mumbai 400005, India}
\date{}

\begin{document}

\maketitle

\begin{abstract}
Many neutron star low-mass X-ray binaries (NS LMXBs) with short orbital periods ($\sim$hours) cycle between outburst and quiescent phases, and thus provide an excellent way to study the accretion process. The cause of such outbursts is believed to be thermal-viscous instability in the accretion disc. However, some of these transient sources show unusually long outbursts. For example, EXO 0748–676 remained in outburst for at least 23 years before entering a quiescence, only to re-emerge 16 years later. We aim to investigate if such long outbursts could be due to the usual disc instability, or if any other mechanism is required. In order to address this question, we systematically compare various properties of long outburst and short outburst NS LMXBs.
For this, we analyze the long-term X-ray light curves of many short orbital period (hours) NS LMXBs, examining the outburst duration and the inferred accretion rate, and estimate the accretion disc mass. Our study shows that long outburst sources are well-separated from the short outburst ones in parameter spaces involving accretion rate, disc mass, outburst duration, etc. in four ways. This implies that the thermal-viscous instability 
in the disc cannot explain the long outbursts, but could explain the short ones. Moreover, we discuss that both donor star related and disc related models have difficulties to explain long outbursts. Our finding will be crucial to understanding the accretion process of transiently accreting neutron stars and black holes.
\end{abstract}

\twocolumn
\section{Introduction}\label{sec:intro}

X-ray outbursts are important probes in detecting transiently accreting neutron star (NS) X-ray binaries (XRBs). Although it is predicted that our Galaxy hosts more than 10$^4$ accreting NS XRBs \citep{kiel2006populating}, they cannot be detected when they are in quiescence. 
Only a fraction of these systems are detected during their outburst when their luminosity increases by a factor of 10$^3$ - 10$^7$ compared to their quiescence luminosity \citep{king1998light}. 
The outburst duration typically lasts for a few weeks and then after a comparatively longer ($\sim$1 yr or more) quiescence duration, they again show another outburst. 
While this is true for most known transient NS low-mass X-ray binaries (LMXBs), a few sources show a much longer outburst duration before transitioning to quiescence again.
Some such sources are EXO 0748--676, 4U 2129+4, 1H 1905+000, etc. \citep{maccarone2022recurrence, heinke2024catalog}. 
For example, EXO 0748--676 remained in outburst for 23 years after its initial detection until 2008, when it entered quiescence \citep{parmar1986discovery,bassa2009optical}. It then emerged into another outburst in June 2024, following 16 years of quiescence \citep{bhattacharya2024xmm}. 
Therefore, the question is what drives these long outbursts.

Historically there were various models proposed to explain the cycle of outburst and quiescence in NS LMXBs. Among them were the disc instability model \citep[DIM; ][]{meyer1981elusive, king1998outbursts, hameury2020review}, the donor star mass-transfer instability
model \citep[MTIM; ][]{bath1975dynamical}, and other models using irradiation \citep{dubus2001disc} or magnetic field \citep{vietri1998new} induced  mass transfer. 
While each model provided insightful physics for specific cases, 
eventually the DIM became the prevailing framework to explain transient accretion as it could be applied to diverse cases of accreting NS or black hole (BH) binaries. 
The DIM attributes the outbursts to thermal-viscous instabilities in the accretion disc \citep{done2007modelling}. 
According to the DIM, if the long-term average accretion rate ($\dot M_{\rm av,long}$) is less than a critical accretion rate ($\dot M_{\rm av,crit}$), 
Initially, the disc temperature and viscosity are small, and the accreted matter piles up as its angular momentum is not sufficiently removed. 
This is a quiescent phase.
As the matter accreted from the donor star piles up, the disc temperature eventually rises above a certain threshold temperature and the disc material (hydrogen) starts getting ionized. This increases the opacity, leading to a thermal instability. Such a  thermal instability causes the viscosity to increase in an unstable manner. 
Thus, the disc material starts losing angular momentum at a higher rate, moves towards the central object, and causes an outburst. 
When most of the disc matter falls onto the central NS or BH, disc temperature decreases, accretion almost stops and a quiescence phase starts again \citep{king1998outbursts, lasota2001disc}.
Naturally, the outburst duration is expected to be determined by the mass content in the disc and the rate of accretion onto the compact object.

Therefore, the long outburst duration of some sources raises important questions about the underlying accretion physics, particularly for NS LMXBs with short orbital periods ($\sim$hours). 
In such a system, a smaller orbital separation suggests a smaller accretion disc with a lower disc mass reservoir, and hence the question is 
whether the standard DIM is still able to explain how these systems can sustain long-duration outbursts, given the limited mass available for accretion. 
In this study, we investigate this and if the long outburst sources are sufficiently different from the short outburst sources.
If they are, then that will open up new directions of research to understand the transient accretion.
Note that understanding the accretion physics in NS and BH X-ray binaries is crucial to probe some aspects of fundamental physics, such as dense matter and strong gravity.
The transient accretion is particularly valuable as it provides access to a broader parameter space and reveals their time evolution, offering more insights beyond what is possible with persistent accretion. 
The current study may open up an excellent opportunity to probe the physics of the accretion process, the binary system, and the donor star for some NS LMXBs. 

This paper is structured as follows. In section~\ref{sec:methods}, we describe both the theoretical calculations done and the observations and data analysis used. 
The results from comparing different NS LMXB sources are presented in section~\ref{sec:results}. 
In section~\ref{sec:disc}, the findings are discussed with the probable alternatives to the disc instability model for long outburst sources.
We present a summary in section~\ref{sec:conc}.

\section{Methods}\label{sec:methods}

In this section, we first select a sample of long and short outbursts 
NS LMXB sources (section~\ref{selection}).
In order to find out if the long outburst sources are a population sufficiently different from short outburst sources, implying a mechanism for long outbursts different from the DIM, we compare these two populations in various parameter spaces involving accretion rates, disc masses, and outburst durations. 
Some of these parameters are theoretically calculated or estimated based on both theoretical calculations and observations (section~\ref{calculations}), while others (e.g., observed outburst duration ($t_{\rm out,obs}$), observed quiescence duration ($t_{\rm quie,obs}$)) are measured from X-ray light curves (section~\ref{observe}). 


\subsection{NS LMXB source selection}\label{selection}

We select NS LMXBs with known binary orbital period ($P_{\rm orb}$) and $M_2/M_1 (=q)$, where $M_2$ and $M_1$ are masses of the donor star and the NS, respectively. 
This is because these parameters are needed for the calculations discussed in section~\ref{calculations}. We select our sample of NS LMXBs from the catalogues:  \citet{heinke2024catalog, salvo2021accretion} (see Tables~\ref{tab:lmxb_orbital}, ~\ref{tab:res_table} in appendix~\ref{app_donor}). 
The long outburst sources we study here have relatively low orbital periods ($0.63-5.24$ hr). Therefore, for a fair comparison, we select short outburst sources also with low orbital periods 
($\sim 0.68 - 21.27$ hr). 
A subset of our short outburst sources are accreting millisecond X-ray pulsars \citep[AMXPs;][]{salvo2021accretion}.

\subsection{Formulae of various parameters}\label{calculations}

In case of the DIM, accretion onto the NS during an outburst happens from the matter piled up in a disc.
The disc mass can either partially or entirely fall onto the NS. We call this fallen disc mass $M_{\rm disc,obs}$, which is $\dot M_{\rm av,out} t_{\rm out,obs}$.
Here, $\dot M_{\rm av,out}$ is the average mass accretion rate during the outburst, which can be estimated from the observed X-ray light curves (see section~\ref{observe}). Thus, $M_{\rm disc,obs}$ can be estimated from observations.

$M_{\rm disc,obs}~(= \dot M_{\rm av,out} t_{\rm out,obs})$ can be less than, equal to, or greater than the entire disc mass, implying accretion of a part of the disc \citep{dubus2001disc}, accretion of the entire disc, or the unsuitability of the DIM, respectively.
In order to find which of these is true, one needs to theoretically calculate the entire disc mass ($M_{\rm disc,calc}$), which can be written as 
\begin{equation}
\label{disc_mass}
    M_{\textrm{disc,calc}} \approx \int_{0}^{R_{\textrm{disc}}}2{\pi}R\Sigma(R)dR,
\end{equation}
where, $\Sigma(R)$ is the surface density as a function of the radial distance $R$, and $R_{\rm disc}$ is the disc outer radius.
Here, the disc inner edge radius is neglected compared to $R_{\rm disc}$.

As this surface density is for the outburst phase, it can be the critical density ($\Sigma_{\textrm{crit}}$) required to trigger an outburst via the thermal-viscous instability \citep[][see also section~\ref{sec:intro}]{cannizzo1988outburst,truss2006decline}. 
This critical density ($\Sigma_{\textrm{crit}}$ in g/cm$^2$), which is assumed to be constant for all disc radii, can be given by
\begin{equation}
\label{sigma}
\Sigma_{\textrm{crit}} =  11.4\alpha_{c}^{-0.86}M_1^{-0.35}R_{\rm disc,10}^{1.05}
\end{equation} 
Here, $M_1$ (in g) is the NS mass, 
$R_{\rm disc,10}$ is the disc outer radius in the unit of  $10^{10}$ cm, and $\alpha_{c}$ is the critical viscosity parameter (threshold where disc transitions from quiescence to outburst).  
Integrating the surface density profile on the radius in equation~\ref{disc_mass}, we get the disc mass (in g) as \citep{truss2006decline}
\begin{equation}
    M_{\textrm{disc,calc}} = 6.04 \times 10^{21}\left(\frac{\alpha_c}{0.02}\right)^{-0.86}\left(\frac{M_1}{1.4}\right)^{-0.35}R_{\textrm{disc,10}}^{3.05}
\end{equation}
Here, $M_1$ is in units of solar mass.
In this paper, we use $\alpha_{c}=0.02$ and $M_1 = 1.4 M_{\odot}$.
A realistic expression of $R_{\rm disc}~(= 1.36 R_{\textrm{circ}})$
was given by \citet{shahbaz1998soft} using the angular momentum conservation. 
Here, the circularization radius ($R_{\textrm{circ}}$) can be written as  \begin{equation}
   \label{Rcirc}
   R_{\textrm{circ}} = a (0.0859q^{-0.426}), 
\end{equation} 
where $a$ is the orbital separation (in cm). We provide the donor mass used to calculate $q$ in Table~\ref{tab:lmxb_orbital} (appendix~\ref{app_donor}). 
In case a range of donor mass values for a given source is available, we use the average value. We use this prescription to estimate the disc outer radius and hence the entire disc mass $M_{\rm disc,calc}$.
The maximum outburst duration ($t_{\rm out,calc}$) is then calculated by
$M_{\rm disc,calc}/\dot M_{\rm av,out}$.

Finally, as mentioned in section~\ref{sec:intro}, a thermal-viscous instability in the disc could happen when $\dot M_{\rm av,long} < \dot M_{\rm av,crit}$.
While, $\dot M_{\rm av,long}$ can be estimated from X-ray observations (section~\ref{observe}), $M_{\rm av,crit}$ (in g/s) could be expressed as
\citep[e.g., ][and references therein]{bhattacharyya2021spin}
\begin{equation}
    \dot{M}_{\textrm{av,crit}} \approx 3.2\times10^{15}\left(\frac{M_1}{M_{\odot}}\right)^{2/3}\left(\frac{P_\textrm{orb}}{3}\right)^{4/3},
    \label{Eq:mdotcrit}
\end{equation}
where both $M_1$ and $M_\odot$ are in g, and orbital period $P_{\rm orb}$ is in hr. This relation can hold if the disc instability is primarily driven by hydrogen ionization and it takes into account the effects of irradiation 
\citep[e.g., ][]{lasota2001disc,dubus2001disc}. 


\subsection{X-ray long-term light curves and accretion rates}\label{observe}

We estimate the durations ($t_{\rm out,obs}$, $t_{\rm quie,obs}$) and accretion rates ($\dot M_{\rm av,out}$, $\dot M_{\rm av,long}$) from long-term X-ray light curves of transient NS LMXBs.
We obtain these long-term light curves of the sources mentioned in Tables~\ref{tab:lmxb_orbital}, \ref{tab:res_table} (appendix~\ref{app_donor}) from three X-ray  monitoring instruments: {\it Rossi X-ray Timing Explorer} ({\it RXTE}) All Sky Monitor (ASM), {\it Monitor of All Sky X-ray Image} ({\it MAXI}), and {\it Swift} Burst Alert Telescope (BAT) 
(details are given in appendix~\ref{long_data}).


We measure 
$t_{\rm out,obs}$ and $t_{\rm quie,obs}$ from light curves as described in appendix~\ref{long_data}. 
We calculate the average accretion rate during outburst ($\dot{M}_{\textrm{av,out}}$) and the long-term average accretion rate ($\dot{M}_{\textrm{av,long}}$) from the corresponding X-ray flux values of sources (flux calculation is described in appendix~\ref{long_data}). 
We use the source distance (see Table~\ref{tab:lmxb_orbital} in appendix~\ref{app_donor}; used the average value for a range of distance) to calculate the luminosity from a flux.
Then, we use an accretion efficiency value (assumed to be 0.1 in this work) to convert a luminosity into an accretion rate.


\begin{figure*}[h!]
\centering
\includegraphics[width=0.95\linewidth, height = 0.7\textwidth]{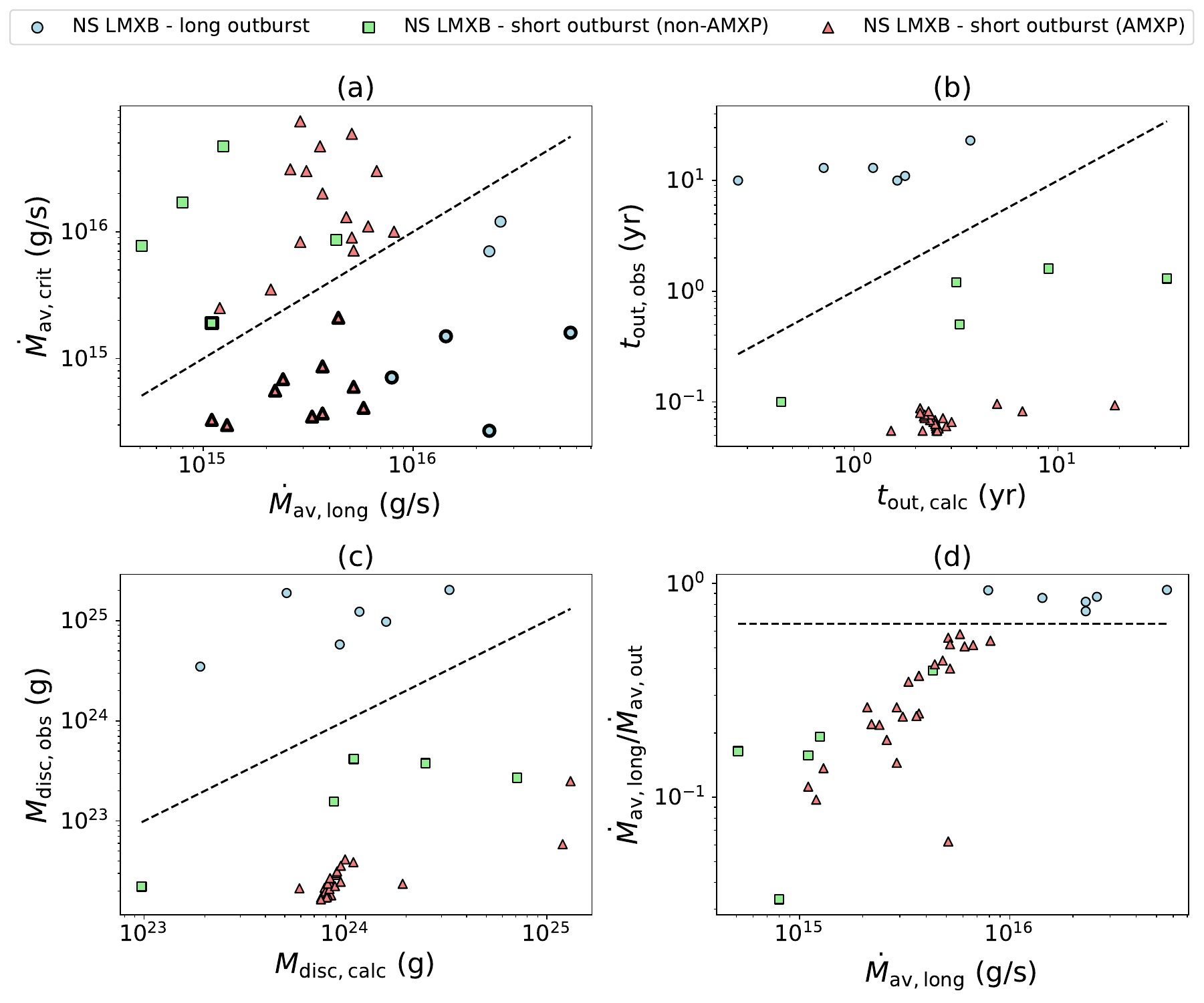}
\caption{Separation of long outburst NS LMXBs and short outburst NS LMXBs (including AMXPs) in four parameter spaces (see sections~\ref{sec:methods} and \ref{sec:results}; Table~\ref{tab:res_table} of appendix~\ref{app_donor}).
{\it Panel (a):} 
Average critical mass transfer rate ($\dot M_{\rm av,crit}$) versus long-term average mass transfer rate ($\dot M_{\rm av,long}$). The 
DIM could explain outbursts for a hydrogen-rich accretion disc above the
$\dot M_{\rm av,crit} = \dot M_{\rm av,long}$ dashed line. 
A source with a symbol marked by a thick black border has a hydrogen-depleted donor star.
{\it Panel (b):} Observed outburst duration ($t_{\rm out,obs}$) versus  calculated maximum outburst duration ($t_{\rm out, calc}$). The $t_{\rm out,obs} = t_{\rm out, calc}$ dashed line separates the short and long outburst sources.
{\it Panel (c):}
Estimated mass of the disc fallen onto the NS during an outburst ($M_{\rm disc,obs}$) and calculated mass in the entire disc ($M_{\rm disc,calc}$). The $M_{\rm disc,obs} = M_{\rm disc,calc}$ dashed line separates the short and long outburst sources.
{\it Panel (d):} 
The ratio of $\dot M_{\rm av,long}$ to estimated average mass accretion rate during outburst ($\dot M_{\rm av,out}$) versus $\dot M_{\rm av,long}$. 
Long outburst sources have higher $\dot M_{\rm av,long}/\dot M_{\rm av,out}$ values, and they are clearly separated from short outburst sources by an  arbitrary horizontal dashed line.}  \label{fig:hist_res}
\end{figure*}

\section{Results}\label{sec:results}

In this paper, we aim to probe if the DIM could explain the outbursts of long outburst NS LMXBs.
In order to achieve this goal, we calculate and estimate various parameter values of a sample of long outburst and short outburst sources, using methods explained in section~\ref{sec:methods}.
Particularly, we analyze many X-ray light curves from these sources to find the durations and accretion rate values. 
These parameter values are presented in Figure~\ref{fig:hist_res}, and Tables~\ref{tab:lmxb_orbital}, \ref{tab:res_table} (appendix~\ref{app_donor}), which compare the properties of long outburst sources with those of short outburst sources (including AMXPs).
Figure~\ref{fig:hist_res} has four panels, each showing a space of one parameter versus another.
As we describe below, each panel  provides a specific insight of physics
from the viewpoint of the DIM. 

Panel (a) of Figure~\ref{fig:hist_res} considers the DIM's basic criterion, viz., $\dot M_{\rm av,long} < \dot M_{\rm av,crit}$, where we consider Eq.~\ref{Eq:mdotcrit} for $\dot M_{\rm av,crit}$ which is for a hydrogen-rich accretion disc.
Here, the $\dot M_{\rm av,crit}$ versus $\dot M_{\rm av,long}$ space is divided with a dashed line corresponding to $\dot M_{\rm av,crit} = \dot M_{\rm av,long}$.
A source with a symbol marked by a thick black
border has a hydrogen-depleted donor star, and hence the above-mentioned DIM's criterion may not entirely apply to them.
But, if we consider other sources with a hydrogen-rich accretion disc,
in spite of a large spread in the $\dot M_{\rm av,crit} - \dot M_{\rm av,long}$ space, 
the short outburst sources (including AMXPs) appear above the $\dot M_{\rm av,crit} = \dot M_{\rm av,long}$ line (i.e., in the space $\dot M_{\rm av,long} < \dot M_{\rm av,crit}$), strongly suggesting that their outbursts can be explained by the DIM.
In the same panel, the long outburst sources 
appear below the dashed line (i.e., in the space $\dot M_{\rm av,long} > \dot M_{\rm av,crit}$), implying their outbursts may not be explained by the DIM. 
Note that $\dot M_{\rm av,long}$ can be a few to tens of times larger than $M_{\rm av,crit}$ for long outburst NS LMXBs,
and similarly smaller than $M_{\rm av,crit}$ for short outburst NS LMXBs with a hydrogen-rich disc (see Table~\ref{tab:res_table} in appendix~\ref{app_donor}). Some of the sources i.e IGR J00291 and Aql X-1, have different reported $\dot M_{\rm av,long}$ in similar work by \cite{coriat2012revisiting}, which can be due to choosing a different timescale or using a different method. This discrepancy doesn't change the overall nature of accretion properties in these three broad categories of sources.  

Figure~\ref{fig:hist_res}(b) presents the observed outburst duration ($t_{\rm out,obs}$) versus the calculated maximum outburst duration ($t_{\rm out,calc}$) for the DIM. 
The DIM could explain an outburst if $t_{\rm out,obs} \le t_{\rm out,calc}$.
While we find a large spread for each of short and long outburst sources, the two populations are cleanly separated.
Moreover, the short outburst NS LMXBs are below the $t_{\rm out,obs} = t_{\rm out,calc}$ line, i.e., in the $t_{\rm out,obs} < t_{\rm out,calc}$ space, and the long outburst NS LMXBs are above the $t_{\rm out,obs} = t_{\rm out,calc}$ line, i.e., in the $t_{\rm out,obs} > t_{\rm out,calc}$ space.
This means the DIM could explain the short outbursts but not the long outbursts.
Typically, $t_{\rm out,obs}$ is about a few to tens of times less than $t_{\rm out,calc}$ for short outburst sources, and similarly greater than $t_{\rm out,calc}$ for long outburst sources
(Table~\ref{tab:res_table} in appendix~\ref{app_donor}).

Figure~\ref{fig:hist_res}(c) is related to Figure~\ref{fig:hist_res}(b), but presented in terms of the disc mass.
This panel presents the estimated mass ($M_{\rm disc,obs}$) of the disc fallen onto the NS during an outburst versus calculated mass in the entire disc ($M_{\rm disc,calc}$).
If $M_{\rm disc,obs}$ is greater than $M_{\rm disc,calc}$, then the DIM cannot explain an outburst because no mass would be available to fall onto the NS if the entire disc is exhausted.
Similar to Figure~\ref{fig:hist_res}(b), here also the long outburst population is cleanly separated from the short outburst population.
The former population is above the $M_{\rm disc,obs} = M_{\rm disc,calc}$ dashed line, implying that the DIM could not explain their outbursts, and the latter population is below the dashed line, implying that the DIM might explain their outbursts.
Table~\ref{tab:res_table} (appendix~\ref{app_donor}) shows that 
typically $M_{\rm disc,obs}$ is a few to tens of times greater than $M_{\rm disc,calc}$ for long outburst sources, and similarly less than $M_{\rm disc,calc}$ for short outburst sources. 
Thus, typically a small fraction (e.g., $\sim 2$\% for Aql X-1) of the disc mass is accreted onto the NS during an outburst for the latter sources.

Finally, Figure~\ref{fig:hist_res}(d) presents $\dot M_{\rm av,long}/\dot M_{\rm av,out}$ for each source. 
Note that the maximum value of this ratio is $1$, which implies a persistent source with a constant accretion rate.
A much lower value of this ratio implies a very low accretion rate in the quiescence period compared to that in the outburst. One expects this for the standard DIM, as the disc cools down after an outburst and the accretion onto the NS becomes very small.
On the other hand, a high value of this ratio implies a level of accretion in the quiescent phase which is not much smaller than that during an outburst. 
The standard DIM cannot possibly explain this and one has to find an extra mechanism of accretion in this case.
Figure~\ref{fig:hist_res}(d) shows that $\dot M_{\rm av,long}/\dot M_{\rm av,out}$ is close to 1 for all long outburst NS LMXBs, and hence the standard DIM may not explain the outbursts of these sources.
On the other hand, while these ratios  for short outburst NS LMXBs (including AMXPs) have a relatively large range ($\sim 0.04-0.6$; Table~\ref{tab:res_table} in appendix~\ref{app_donor}), all of them are lower than, and hence cleanly separated from, the values for long outburst sources.

Thus, in order to check if the DIM could explain the outbursts of long outburst NS LMXBs, we look into three aspects of the DIM, viz., 
(i) $\dot M_{\rm av,long} < \dot M_{\rm av,crit}$ (hydrogen-rich disc instability criterion);
(ii) $M_{\rm disc,obs} \le M_{\rm disc,calc}$ 
(availability of sufficient disc mass to make the observed outburst possible); and 
(iii) $\dot M_{\rm av,long}/\dot M_{\rm av,out}$ is less than and not close to $1$ (criterion of cooled down disc making the accretion level very low during quiescence).
For long outburst NS LMXBs, these criteria are not satisfied, and hence we conclude that the standard DIM cannot explain their outbursts.
Considering these criteria, the short outburst NS LMXBs are somewhat cleanly separated from the long outburst NS LMXBs, and we find that the DIM could explain the outbursts of the former.

\section{Discussion}\label{sec:disc}

In this work, we find that the accretion mechanism in NS LMXBs with long outbursts should be different from that in short outburst LMXBs (including AMXPs). 
Particularly, the standard disc instability model (DIM) 
is not sufficient to explain outbursts of the former sources, while this model could explain the outbursts of the latter sources.
Then, what is the origin of alternate long outbursts and quiescence periods?
These could be regulated either by the donor star or by the accretion disc, and the most important question is which one.
Here, we briefly mention some of the plausible mechanisms (see also section~\ref{sec:intro}), and the challenges to explain the transient phenomena of long outburst NS LMXBs. 

The theory of donor star instability \citep{bath1975dynamical} could be revived to explain long-duration outbursts in transient NS LMXBs. 
The donor star's envelope may experience thermal instability and that could make it expand further into the NS Roche lobe increasing the mass transfer \citep[mass-transfer instability model (MTIM); ][]{bath1975dynamical}.
Also, either thermal or magnetic processes \citep{ritter2008irradiation,zhu2012donors} in the donor star or irradiation-driven mass transfer could change in donor mass loss rate, and hence cause the alternate outburst and quiescence phases of long outburst sources.
However, there are varieties of donor stars in our sample of long outburst sources (Table~\ref{tab:lmxb_orbital} in appendix~\ref{app_donor}), such as white dwarf, subgiant and M-dwarf.
Therefore, perhaps a single instability model cannot work for all donors of the long outburst NS LMXBs.

The accretion disc,
with a modification of the standard DIM, might explain 
the long outbursts. 
Some of such models involve 
irradiation controlled accretion \citep{dubus2001disc}, 
angular momentum transport \citep{echeveste2024impact}, 
non-standard viscosity distribution \citep{merloni2006limit}, etc. 
For example, the accretion rate might be modified by irradiation on the disc \citep{dubus2001disc}. In this case, the outer layer is kept ionized by the X-ray radiation. 
This prevents the disc from cooling down, and in turn, the accretion continues for longer than the predicted duration by standard DIM. 
In the case of the enhanced angular momentum transport, the material is transported outwards due to magneto-rotational instability and turbulent viscosity \citep{kulkarni2008accretion, balbus1990powerful}.
This causes a redistribution of matter within the disc that can prevent the propagation of a cold front and keep the outer disc hot and ionized sustaining the accretion. 
Some other mechanisms involve warped or precessing disc \citep{ogilvie2001precessing}, magnetic field stabilized accretion \citep{skadowski2016magnetic}, and tidal torques or resonances due to the donor star and disc interaction \citep{hameury2020review}. 
However, none of the above disc-related models can possibly work if the disc does not have sufficient mass to sustain a long outburst.
Indeed, we find that the disc mass is a few to tens of times lower than required for long outbursts NS LMXBs (see section~\ref{sec:results}; Table~\ref{tab:res_table} of appendix~\ref{app_donor}).

A solution to this problem could be for the disc to have more mass than we estimate here.
This, along with the above-mentioned mechanisms, such as the one which keeps the outer disc sufficiently hot to make some level of accretion possible after the outburst, could explain why the quiescent phases of long outburst sources are not very faint.
However, such a high disc mass might imply that the disc surface density ($\Sigma$) is a few to tens of times greater than that estimated from Eq.~\ref{sigma}
(see section~\ref{sec:results}; Table~\ref{tab:res_table} of appendix~\ref{app_donor}).
Even if we consider a hydrogen-depleted disc, it could be challenging to explain such a high $\Sigma$, particularly for a few selected NS LMXBs which show long outbursts.
Besides, not all such long outburst sources have a hydrogen-depleted disc.
Therefore, our results point to the necessity of further detailed theoretical studies to understand the transient accretion mechanism of NS and BH X-ray binaries.

\section{Summary}\label{sec:conc}

This study suggests a significant diversity in accretion behaviours across different classes of transient LMXBs.
The standard thermal-viscous disc instability model could explain the accretion in short outburst NS LMXBs, but not that in long outburst ones.
We show that both the donor star related models and the accretion disc related models have difficulties in explaining long outbursts.


    


\bibliography{references}{}
\bibliographystyle{elsarticle-harv}

\appendix

\section{Source properties and donor star mass}\label{app_donor}
\setcounter{table}{0}
\setcounter{figure}{0}
\renewcommand{\thetable}{A\arabic{table}}
\renewcommand{\thefigure}{A\arabic{figure}}
In this section, we present our sample (section~\ref{selection}) of long outburst NS LMXBs and short outburst NS LMXBs (including AMXPs) with their binary orbital period, distance and donor star mass (Table~\ref{tab:lmxb_orbital}).
We also present the calculated, estimated and observed source parameters in Table.~\ref{tab:res_table} (see also section~\ref{sec:methods} and appendix~\ref{long_data}).

\begin{table*}
    \centering
    \small
    \caption{Properties of a sample of transient NS LMXBs (appendix~\ref{app_donor}; see also section~\ref{sec:methods}).} \vspace{0.5cm}
    \resizebox{1.15\textwidth}{!}{%
    \begin{tabular}{cccccc}
        \hline
        Source  & Orbital period & Distance & Donor mass & References \\
        Name & (hr) & (kpc) & ($M_\odot$) & \\
        
        \hline
        \multicolumn{5}{c}{LMXBs with long outburst} \\ \hline
        IGR J17062--6143       & 0.63  & 6.8--7.8 & 0.01 & 1 \\
        1M 1716--315           & 1.00  & 6.9 & 0.01 & 2  \\
        1H 1905+000            & 1.33  & 7--10 & 0.02 & 3, 4 \\
        HETE J1900--2455       & 1.39  & 4.3 & 0.05 & 5  \\
        EXO 0748--676          & 3.82  & 5.9 -- 7.7 & 0.11--0.15 & 6 \\
        4U 2129+47             & 5.24  & 6.3 & 0.4--0.8 & 7 \\
        \hline
        \multicolumn{5}{c}{LMXBs with short outburst} \\ \hline
        XMM J174457--2850.3    & 1.7   & 6.5 & 0.1 & 8  \\
        Swift J1922.7--1716    & 5.0   & 5--11 & 0.15 & 9 \\
        AX J1754.2--2754       & 5.4   & 6.3--9.6 & 0.2 & 10 \\
        AX J1745.6--2901       & 8.35  & 6--10 & 0.4 & 11 \\
        MAXI J0556--332        & 16.41 & 43.6 & 0.07 & 12 \\
        \hline
        \multicolumn{5}{c}{LMXBs with short outburst (AMXPs)} \\ \hline
        XTE J1807--294         & 0.68  & 5.5 & 0.02 & 13 \\
        XTE J1751--305         & 0.71  & 11 & 0.014 & 14 \\
        XTE J0929--314         & 0.73  & 7.4 & 0.008 & 15 \\
        MAXI J0911--655        & 0.74  & 9.5 & 0.05 & 16 \\
        IGR J16597--3704       & 0.77  & 9.1 & 0.02 & 16 \\
        Swift J1756.9-2508     & 0.91  & 8.5 & 0.03 & 17 \\
        NGC 6440 X-2           & 0.96  & 8.3 & 0.01 & 18 \\
        MAXI J1957+032         & 1.16  & 3--7 & 0.018 & 19 \\
        IGR J17494--3030       & 1.25  & 8 & 0.02 & 20 \\
        IGR J17379--3747       & 1.88  & 8.5 & 0.06 & 21 \\
        SAX J1808.4--3658      & 2.01  & 3.5 & 0.05 & 22 \\
        IGR J00291+5934        & 2.46  & 3.7--4.7 & 0.1 & 23 \\
        IGR J17511--3057       & 3.47  & 3.6 & 0.02 & 24 \\
        IGR J17498--2921       & 3.84  & 7.6 & 0.17 & 25 \\
        XTE J1814--338         & 4.27  & 9.6 & 0.2--0.3 & 26 \\
        PSR J1023+0038         & 4.75  & 1.4 & 0.2 & 27 \\
        MAXI J1816--195        & 4.83  & 6.3 & 0.1--0.5 & 28 \\
        SRGA J144459.2--604207 & 5.22  & 8--9 & 0.2 & 29 \\
        XSS J12270--4859       & 6.91  & 1.4--3.6 & 0.1--0.3 &  30 \\
        SAX J1748.9--2021      & 8.76  & 8.5 & 0.7--0.8 & 31 \\
        Swift J1749.4--2807    & 8.86  & 5.4--8 & 0.7--1.0 & 32 \\
        IGR J17591--2342       & 8.80  & 7.6 & 0.3--0.5 & 33 \\
        IGR J18245--2452       & 11.03 & 5.5 & 0.6 & 34 \\
        Aql X-1                & 18.95 & 5.75 & 0.4--0.8 & 35 \\
        IGR J17480--2446       & 21.27 & 5.9 & 0.13 &  36\\\hline\\
        
        \multicolumn{5}{l}{1. \citet{hernandez2019multiwavelength}, 2. \citet{jonker2007quasi}, 3. \citet{jonker2007cold}, 4. \citet{jonker2006neutron}}\\
        \multicolumn{5}{l}{5. \citet{elebert2008optical}, 6. \citet{hynes2009quiescent}, 7. \citet{bothwell2008spectroscopic}, 8. \citet{degenaar2014peculiar}}\\
        \multicolumn{5}{l}{9. \citet{degenaar2012two}, 10. \citet{shaw2017near}, 11. \citet{ponti2018nustar+}, 12. \citet{cornelisse2012nature}}\\
         \multicolumn{5}{l}{13. \citet{chou2008precise}, 14. \citet{jonker2003search}, 15. \citep{jonker2004optical}, 16. \citet{marino2019indications}}\\
         \multicolumn{5}{l}{17. \citet{sanna2018swift}, 18. \citet{heinke2010discovery}, 19. \citet{sanna2022maxi}, 20. \citet{ng2021nicer}}\\
         \multicolumn{5}{l}{21. \citet{sanna2018xmm}, 22. \citet{deloye2008optical}, 23. \citet{jonker2008optical}}\\
         \multicolumn{5}{l}{24. \citet{bozzo2010swift}, 25. \citet{galloway2024inferring}, 26. \citet{baglio2013long}, 27. \citet{papitto2018first}}\\
         \multicolumn{5}{l}{28. \citet{bult2022discovery}, 29. \citet{ng2024nicer}, 30. \citet{de2015multiwavelength}, 31. \citep{cadelano2017optical}}\\
         \multicolumn{5}{l}{32. \citet{degenaar2012quiescent}, 33. \citet{sanna2018nustar}, 34. \citet{de2017transitional}, 35. \citet{mata2017donor}, 36. \citet{testa2012near}}
        \\ 
    \end{tabular}
    }
    \label{tab:lmxb_orbital}
\end{table*}

\begin{table*}[h!]
\centering
\caption{\centering 
Observed, estimated, and 
calculated parameters for transient NS LMXBs, including AMXPs (appendix~\ref{app_donor}; see also sections~\ref{sec:methods} and \ref{sec:results}).}
\label{tab:res_table}
\resizebox{\textwidth}{!}{%
\begin{tabular}{c c c c c c c c c}

\hline
    Source &  $P_{\textrm{orb}}$\footnotemark[1] & $t_{\textrm{out,obs}}$\footnotemark[2] & $\dot{M}_{\textrm{av,out}}$\footnotemark[3] & $t_{\textrm{out,calc}}$\footnotemark[4] & $M_{\textrm{disc,obs}}$\footnotemark[5]& $M_{\textrm{disc,calc}}$\footnotemark[6]& $\dot{M}_{\textrm{av,crit}}$\footnotemark[7] & $\dot{M}_{\textrm{av,long}}$\footnotemark[8]\footnotetext[1]{Orbital period (hours).}\\
    Name &  (hr)  &  & (g/s) & (yr) & (g) & (g) & (g/s) & (g/s) \\
    \hline
    \multicolumn{8}{c}{LMXBs with long outburst}   \\ \hline
     IGR J17062--6143 & 0.63 & 10 yr & 3.1E16 & 1.63  & 9.77E24 & 1.6E24 & 4.9E14 & 2.3E16\\
     1M 1716--315 & 1.0 & 13 yr & 8.5E15 & 0.71  & 3.48E24 & 1.89E23 & 9.6E14 & 7.9E15\\
     1H 1905+000 & 1.33 & 11 yr & 1.67E16 & 1.78  & 5.79E24 & 9.17E23  & 1.5E15 & 1.43E16\\
     HETE J1900--2455 & 1.39 & 10 yr & 6.0E16 & 0.27 & 1.89E25 & 5.13E23 & 1.6E15 & 5.6E16\\
     EXO 0748--676 & 3.82 & 23 yr & 2.8E16 & 3.72  &  2.03E25 & 3.29E24 &  5.78E15 & 2.3E16 \\ 
     4U 2129+47 & 5.24 & 13 yr & 3.0E16 & 1.24 & 1.23E25 & 1.22E24 & 8.8E15 & 2.6E16\\
     
     \hline
    \multicolumn{8}{c}{LMXBs with short outburst}   \\ \hline
     XMM J174457--2850.3 & 1.7 & 0.1 yr & 7E15 & 0.44 & 2.21E22 & 9.8E22 & 1.9E15 & 1.1E15\\
     Swift J1922.7--1716 & 5.0 & 1.6 yr & 3.1E15 & 9.01 & 1.56E23 & 8.8E23 & 8.3E15 & 5.1E14\\
     AX J1754.2--2754 & 5.4 & 1.2 yr & 1.1E16 & 3.17  & 4.16E23 & 1.1E24 & 9.2E15 & 4.3E15 \\
     AX J1745.6--2901 & 8.35 & 0.5 yr & 2.4E16 & 3.3 &  3.78E23 & 2.5E24 & 1.7E16 & 0.8E15\\
      MAXI J0556--332 & 16.41 & 1.3 yr & 6.5E15 & 34.26 &  2.7E23 & 7.0E24 & 4.7E16 & 1.25E15\\
     \hline
     \multicolumn{8}{c}{LMXBs with short outburst (AMXPs)} \\ \hline
     XTE J1807-294    & 0.68  & 21 d & 9.5E15   & 2.53  & 1.72E22  & 7.60E23 & 5E14  & 1.3E15\\
     XTE J1751--305    & 0.71  & 22 d & 9.8E15   & 2.52 & 1.86E22  & 7.80E23 & 6.1E14  & 1.1E15\\
     XTE J0929--314    & 0.73  & 22 d & 9.5E15   & 2.83   & 1.81E22  & 8.50E23 & 6.3E14  &  6.3E15\\
     MAXI J0911--655   & 0.74  & 25 d & 1.0E16   & 2.5   & 2.16E22  & 8.00E23 & 6.4E14 & 3.7E15\\
     IGR J16597--3704  & 0.77  & 21 d & 1.0E16   & 2.6  & 1.81E22  & 8.20E23 & 6.8E14  & 6.8E15\\
     Swift J1756.9--2508  & 0.91  & 23 d & 1.0E16   & 2.53  & 1.98E22  & 8.00E23 & 8.5E14  & 5.2E15 \\
     NGC 6440 X-2      & 0.96  & 24 d & 1.1E16   & 2.45  & 2.28E22  & 8.50E23 & 8.9E14  & 2.4E15\\
     MAXI J1957+032  & 1.16  & 20 d & 1.0E16   & 2.54 & 1.72E22  & 8.00E23 & 1.1E14  & 2.2E15\\
     IGR J17494--3030  & 1.25  & 20 d & 1.0E16   & 2.56  & 1.73E22  & 8.10E23 & 1.3E15  &  3.7E15\\
     IGR J17379--3747  & 1.88  & 23 d & 1.05E16  & 2.5  & 2.1E22  & 8.30E23 & 2.2E15  & 4.4E15\\
     SAX J1808.4--3658 & 2.01  & 20 d & 1.23E16  & 1.52  & 2.13E22  & 5.90E23 & 2.5E15  & 1.2E15\\
     IGR J00291+5934  & 2.46  & 24 d & 7.96E15  & 3.01  & 1.65E22  & 7.57E23 & 3.21E15  & 2.1E15 \\
     IGR J17511--3057  & 3.47  & 30 d & 9.14E15  & 6.7   & 2.36E22  & 1.95E24 & 5E15  & 3.1E15\\
     IGR J17498--2921  & 3.84  & 20 d & 1.3E16   & 2.17  & 2.24E22  & 8.9E23 & 5.82E15  & 2.2E15\\
     XTE J1814--338 & 4.27  & 25 d & 1.1E16   & 2.36  & 2.37E22  & 8.20E23 & 6.7E15  & 2.9E15\\
     PSR J1023+0038   & 4.75  & 28 d & 1.2E16   & 2.37  & 2.90E22  & 9.00E23 & 7.7E15  & 2.9E15\\
     MAXI J1816--195 & 4.83  & 26 d & 1.2E16   & 2.22 & 2.69E22  & 8.40E23 & 7.91E16  & 6.1E15\\
     SRGA J144459.2--604207 & 5.22  & 26 d & 1.1E16   & 2.73  & 2.47E22  & 9.50E23 & 8.8E15  & 4.8E15\\
     XSS J12270--4859  & 6.91  & 32 d & 1.5E16   & 2.11 & 4.14E22  & 1.00E24 & 1.2E16  & 4.7E15\\
     SAX J1748.9--2021   & 8.76  & 27 d & 1.3E16   & 2.2  & 3.03E22  & 9.00E23 & 1.75E16 & 3.1E15\\
     IGR J17591--2342 & 8.80  & 28 d & 1.3E16   & 2.22 & 3.14E22  & 9.10E23 & 1.76E16  & 6.7E15\\
     Swift J1749.4--2807  & 8.86  & 29 d & 1.4E16   & 2.11  & 3.57E22  & 9.30E23 & 1.78E16  & 2.6E15\\
     IGR J18245--2452 & 11.03 & 30 d & 1.5E16   & 2.32  & 3.88E22  & 1.10E24 & 2.3E16 & 3.6E15\\
     Aql X-1 & 18.95 & 35 d & 8.2E16   & 5.03  & 2.5E23   & 1.3E25  & 4.9E16  & 5.1E15\\ 
     IGR J17480--2446  & 21.27 & 34 d & 2.0E16   & 19.02  & 5.87E22  & 1.20E25 & 5.7E16  & 2.9E15\\ \hline \\
     
     \multicolumn{4}{l}{\small $^1$Binary orbital period.} \\
     \multicolumn{4}{l}{\small $^2$Observed outburst duration.} \\
     \multicolumn{8}{l}{\small $^3$Estimated average mass accretion rate during outburst.} \\
     \multicolumn{8}{l}{\small $^4$Calculated maximum duration of outburst (if the entire disc mass falls onto the NS).} \\
     \multicolumn{8}{l}{\small $^5$Estimated mass of the disc fallen onto the NS during an outburst (entire disc mass may or may not fall).} \\    
     \multicolumn{8}{l}{\small $^6$Calculated mass in the entire disc.} \\
     \multicolumn{8}{l}{\small $^7$Average critical mass transfer rate.} \\
     \multicolumn{8}{l}{\small $^8$Long-term average mass transfer rate.} \\
        
\end{tabular}%
    }
\end{table*}

\section{Long-term light curve data analysis
\label{long_data}}
\setcounter{table}{0}
\setcounter{figure}{0}
\renewcommand{\thetable}{B\arabic{table}}
\renewcommand{\thefigure}{B\arabic{figure}}
\begin{figure*}[ht]
    \centering
\includegraphics[width=0.98\linewidth]{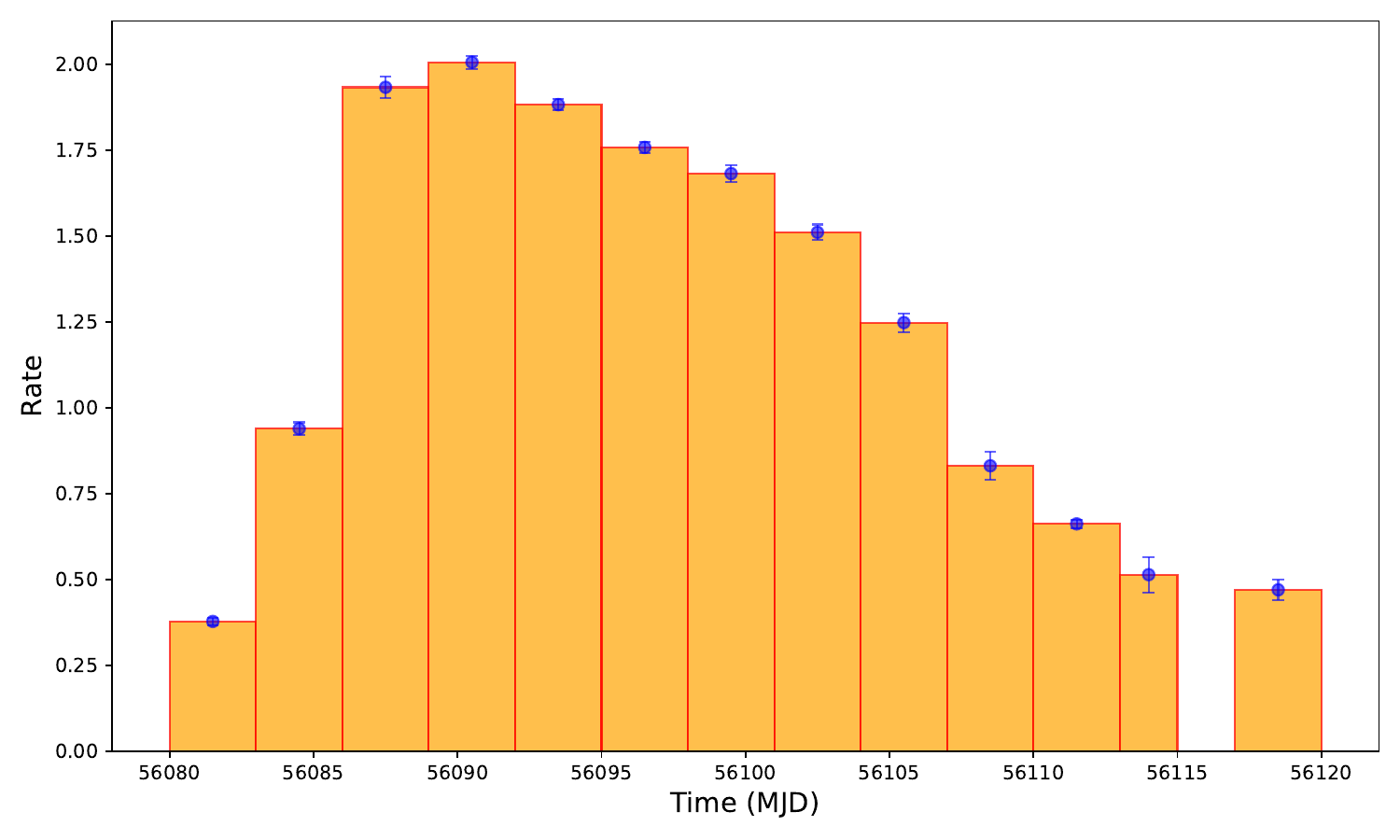}
\caption{Long-term {\it MAXI} light curve of Aql X-1 as an example to explain the method described in section~\ref{sec:methods} and Appendix~\ref{long_data}. 
The width of each bar is typically three days, and the height represents the average count rate for a three-day bin.}
\label{fig:long_lc}
\end{figure*}

Here, we provide a description of the analysis of the long-term X-ray light curves from the sources listed in Table~\ref{tab:lmxb_orbital}, \ref{tab:res_table} (appendix~\ref{app_donor}). 
The light curve data were extracted from three X-ray  monitoring instruments: {\it Rossi X-ray Timing Explorer} ({\it RXTE}) All Sky Monitor (ASM), {\it Monitor of All Sky X-ray Image} ({\it MAXI}), and {\it Swift} Burst Alert Telescope (BAT). 
These instruments cover different energy ranges and have different sensitivities, which influence our choice of data based on the availability and energy coverage of the source observations.\\
\textbf{\textit{RXTE}-ASM:} This instrument provided long-term monitoring data of X-ray sources from 1996 until its decommissioning in 2012. 
The {\it RXTE}-ASM data is preferred to study outbursts that occurred before 2009, when {\it MAXI} data were not available. 
The energy range of 2--12 keV captures soft X-ray emission, which is crucial for detecting the outburst behaviour of NS LMXBs. 
For ASM, 1 Crab corresponds to 75 counts/s.\\
\textbf{\textit{MAXI}: } The Monitor of All-Sky X-ray Image (MAXI) onboard the International Space Station has been monitoring the X-ray sky since 2009. {\it MAXI}’s broader energy range (2--20 keV) allows one to track both soft and moderately hard components of X-ray emissions. For observations after 2009, {\it MAXI} data are utilized preferentially due to its better sensitivity in the soft X-ray band. For MAXI, 1 Crab corresponds to 1.6 photons/cm$^2$/s.\\
\textbf{\textit{Swift}-BAT:} The BAT on the {\it Swift} satellite is optimized for hard X-ray (15--50 keV) observations. Given the harder energy coverage, BAT is less sensitive to the thermal emission from the accretion disc but can capture non-thermal emission, e.g., from coronal activities. 
For BAT, 1 Crab corresponds to 0.22 counts/cm$^2$/s.  

{\bf Outburst duration calculation: }
The outburst durations for non-AMXP NS LMXBs (with both long and short outbursts) are reported from a recent catalogue of neutron star outbursts \citep{heinke2024catalog}. For AMXPs, there exists various reported outburst durations in the literature, and often not well constrained. Hence, we calculated that for each source, we considered the last three outbursts when more than three are available and obtained the average light curve. If not, we use the latest outbursts to get an average light curve. The light curve is then modeled using a linear rise followed by an exponential decay. First, a time is identified from the profile, which is considered the rising time, and then a linear fit till the maximum counts in the profile determines the time of rise. Then from the maximum point, an exponential decay curve is fitted to the rest of the light curve of the form: $Ae^{(-(t-\tau))}$. Then the decay time ($\tau$) from the fit is added to the rise time to get the total duration. The quiescence duration is the average difference between the last three outbursts (when available). If the last outburst is continuing, then the quiescence duration till October 2024 is considered.

{\bf Flux Calculation:} 
In order to enhance the signal-to-noise ratio and to avoid short-term variability, we bin the non-AMXP LMXB light curves to a weekly average and the AMXP light curves to a three-day average (e.g., Fig~\ref{fig:long_lc}). We get the total count rate by adding the counts in each bin of the light curve for both outburst and total (outburst+quiescence) duration (see Fig~\ref{fig:long_lc}). Then the total count rate is divided by the respective durations to get the average rate. For each instrument, the average count rates are converted to average flux using the relation between count rates and the Crab flux for that particular instrument.

\end{document}